\documentclass[%
 reprint,
 amsmath,amssymb,
 aps,
prb,
]{revtex4-2}

\usepackage{graphicx}
\usepackage{dcolumn}
\usepackage{bm}
\usepackage{xcolor}
\definecolor{vastkust}{RGB}{0, 48, 80} 
\usepackage[hidelinks]{hyperref}
\usepackage{url}
\hypersetup{
  colorlinks,
  citecolor=vastkust,
  linkcolor=vastkust,
  urlcolor=vastkust}

\begin{document}

\preprint{APS/123-QED}

\title{Mott insulating negative thermal expansion perovskite TiF$_3$}

\author{Donal Sheets}
\affiliation{Department of Physics, University of Connecticut, Storrs, Connecticut, 06269 USA}
\affiliation{Institute for Materials Science, University of Connecticut, Storrs, Connecticut, 06269 USA}
\author{Kaitlin Lyszak}
\affiliation{Department of Physics, University of Connecticut, Storrs, Connecticut, 06269 USA}
\affiliation{Institute for Materials Science, University of Connecticut, Storrs, Connecticut, 06269 USA}
\author{Menka Jain}
\affiliation{Department of Physics, University of Connecticut, Storrs, Connecticut, 06269 USA}
\affiliation{Institute for Materials Science, University of Connecticut, Storrs, Connecticut, 06269 USA}
\author{Gayanath W. Fernando}
\affiliation{Department of Physics, University of Connecticut, Storrs, Connecticut, 06269 USA}
\affiliation{Institute for Materials Science, University of Connecticut, Storrs, Connecticut, 06269 USA}
\author{Ilya Sochnikov, Jacob Franklin}
\affiliation{Department of Physics, University of Connecticut, Storrs, Connecticut, 06269 USA}
\affiliation{Institute for Materials Science, University of Connecticut, Storrs, Connecticut, 06269 USA}
\author{Jason N. Hancock}
\affiliation{Department of Physics, University of Connecticut, Storrs, Connecticut, 06269 USA}
\affiliation{Institute for Materials Science, University of Connecticut, Storrs, Connecticut, 06269 USA}
\author{R.~Matthias Geilhufe}
\affiliation{Department of Physics, Chalmers University of Technology, 412 96 G\"{o}teborg, Sweden}

\date{\today}

\begin{abstract}
We characterize perovskite TiF$_3$, a material which displays significant negative thermal expansion at elevated temperatures above its cubic-to-rhombohedral structural phase transition at 330 K. We find the optical response favors an insulating state in both structural phases, which we show can be produced in density functional theory calculations only through the introduction of an on-site Coulomb repulsion. Analysis of the magnetic susceptibility data gives a $S$=$\frac{1}{2}$ local moment per Ti$^{+3}$ ion and an antiferromagnetic exchange coupling. Together, these results show that TiF$_3$ is a strongly correlated electron system, a fact which constrains possible mechanisms of strong negative thermal expansion in the Sc$_{1-x}$Ti$_x$F$_3$ system. We consider the relative strength of the Jahn-Teller and electric dipole interactions in driving the structural transition.
\end{abstract}

\maketitle

\section{\label{sec:introduction}Introduction}
Materials based on the perovskite lattice structure have garnered sustained interest for over a century, largely because its accommodation of ionic substitutions permits a wide range of effective controls with which one can explore quantum matter. Along with its natural cascade of structural phases, the perovskite lattice produces opportunities to promote phase competition and pursue exotic behavior as demonstrated with the discovery of colossal magnetoresistance\cite{Ramirez1997}, high-$\kappa$ dielectricity\cite{Song2022}, strong negative thermal expansion (NTE)\cite{Rodriguez2009,Greve2010}, and high-temperature superconductivity\cite{Bednorz1986}. Within the context of these discoveries, technological advances such as ion-shuttling\cite{Sun2018} and solar harvesting\cite{Pellet2014} relevant to energy applications have been demonstrated. 
The transition metal fluoride perovskites are much less studied than their oxide cousins, and display marked differences. At the level of considering whether chemically stable systems could exist, oxide perovskites have a typical form ABO$_3$, where the nominal valence sum of the A and B sites is near +6. There are very few examples of A-site free oxide perovskites, ReO$_3$ and MoO$_3$ among them\cite{Rodriguez2009}, consistent with the rarity of hexavalent single ions. The fluoride perovskites however, with typical formula unit BF$_3$, require only a trivalent B site, which is common across the $d$-filling transition metal series (B=Sc,Ti,V,Cr,Mn,Fe,Co,Ni)\cite{MogusMilankovi1985,Lee2018,Kennedy2002,Kennedy2004,Greve2010}. 

\begin{figure}
	\includegraphics[width=3. in]{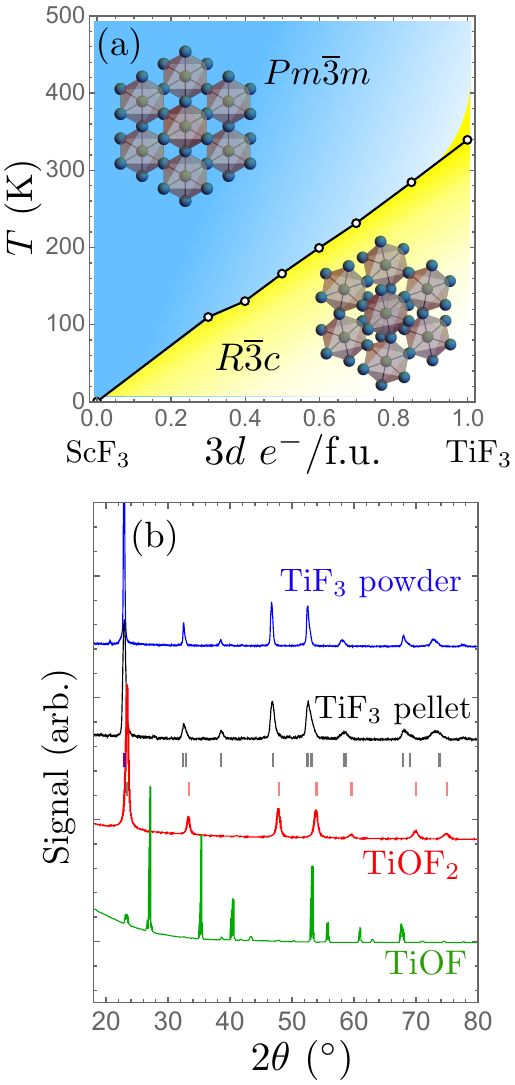}
		
	\caption{(a) Structural phase diagram of Sc$_{1-x}$Ti$_x$F$_3$. The horizontal axis is the $x$ in this composition, which also corresponds to the expected 3$d$ filling in this system. Structural data are from refs. \cite{Morelock2015} and \cite{Handunkanda2015}. (b) X-ray diffraction data on loose TiF$_3$ powder (blue), pelletized TiF$_3$ powder (black), and for comparison X-ray diffraction data on possible oxidation products TiOF$_2$ ($Pm\overline{3}m$ structure, ref. \cite{Powell2015}) and TiOF ($P4_2/mnm$, ref. \cite{Cumby2018}). No peaks from either oxide are present in the samples presented here. Vertical bars show the expected diffraction peaks for TiF$_3$ (black) and TiOF$_2$ (red).}
	\label{fig:structure}
\end{figure}

Among the many potential BF$_3$ materials based on 3$d$-filling transition metal B ions, the end member ScF$_3$ is particularly simple electronically and magnetically. Considering Pauling valence counting returns a $d^0$ configuration for this ion, which is fully consistent with the large electronic band gap both observed via resonant photoemission\cite{Umeda1996} and computed using density functional theory methods\cite{Bocharov2016} as well as its diamagnetic response\cite{Hu2014}. There is however a rich and robust physical peculiarity in this system, in particular the sizable \textit{negative} thermal expansion coefficient ($\alpha_L\geq$-14 ppm/K) present over a 1000 $K$ temperature window\cite{Greve2010}. Soft mode spectroscopy using inelastic X-ray scattering has determined that the ground state of this system resides very close to a structural quantum phase transition between the cubic highly symmetric $Pm\overline{3}m$ and a lowered-symmetry cell-doubled rhombohedral $R\overline{3}c$ structure\cite{Handunkanda2015}. It is widely believed that the associated strong transverse fluctuations of the F ions are responsible for the NTE in ScF$_3$\cite{WANG2014,Wendt2019,Li2011}, although the degree of atomic correlation within the unit cell is debated\cite{Wei2020,Zaliznyak2021}. Research into the anomalous NTE of trifluorides is a very active area\cite{Greve2010,Morelock2013,Morelock2014,Morelock2015,Handunkanda2015,Handunkanda2016,Handunkanda2019,Cumby2018,Hu2014,WANG2014,Wendt2019,Li2011,Wei2020,Zaliznyak2021,Occhialini2018,Corrales-Salazar2017,Bocharov2016,Purans2016,Piskunov2016,Qin2022}, since ScF$_3$ displays behavior typical of a broader class of more complex systems such as ZrW$_2$O$_8$ and related materials which also show very soft phonon modes and display strong negative thermal expansion\cite{Mary1996,Chen2015,Shi2021}.


Perturbations of the structural quantum phase transition and strong NTE behavior using ionic substitution on the trivalent Sc$^{+3}$ site in Sc$_{1-x}$X$_{x}$F$_3$ has revealed the NTE behavior to be persistent at elevated temperatures for X=Al,Y,Ti,Ga,Fe over a significant span of compositions\cite{Morelock2013,Morelock2014,Morelock2015,Hu2014}. The opportunity to compositionally introduce electronic degrees of freedom in a system with a structurally critical ground state is an exciting prospect, particularly in light of the high interest in low-carrier superconductivity in elemental bismuth\cite{Prakash2017} as well as incipient-ferroelectric perovskites SrTiO$_3$\cite{Gastiasoro2020} and KTaO$_3$\cite{Liu2021}. The substitution X=Ti in particular permits ionic substitution over its entire compositional range\cite{Morelock2014} and appears to accordingly introduce muted disorder effects\cite{Occhialini2018} in comparison to other substitutions (X=Al,Y). Figure \ref{fig:structure}a summarizes the known phase diagram of Sc$_{1-x}$Ti$_{x}$F$_3$, showing a structurally lowered $R\overline{3}c$ symmetry for all temperatures below its structural transition temperature $T_s(x)\simeq 340K \times x$. 
This phase boundary along with the documented pressure behavior has been analyzed using a Landau-Ginzberg-Devonshire approach to identify some compositions in this series with strong promise for enabling near-ambient barocaloric refrigeration\cite{Corrales-Salazar2017}. Of basic interest is whether doping ScF$_3$ with electrons to produce TiF$_3$ results in a metallic or insulating state, and whether there is any long range magnetic order resultant from spin exchange interactions. At least two distinct possibilities have been presented in a prior theoretical work\cite{Perebeinos2004}: a ferromagnetic half metal in the uncorrelated ($U=0$) limit where the calculations are well controlled, versus incorporating electronic correlation ($U\simeq 8$ eV) predicts an antiferromagnetic insulator with spin-$\frac{1}{2}$ moments. In this work, we constrain the electronic and magnetic status of TiF$_3$ in a combined study of its structural, magnetic, and electronic properties\cite{Perebeinos2004}.

\section{\label{sec:Structural}Structural characterization}

Powder samples of TiF$_3$ sourced from Sigma-Aldrich were used in this study and phase purity was confirmed using X-ray diffraction. Figure \ref{fig:structure}b shows room temperature powder diffraction scans from a pressed pellet and loose powder collected from a Bruker D2 Phaser diffractometer and a Cu X-ray source. Also shown are the expected Bragg peak positions for potential oxide impurities TiOF\cite{Cumby2018} and TiOF$_2$\cite{Powell2015}. 
All observed peaks can be assigned to the expected $R\overline{3}c$ Bragg reflections of TiF$_3$, confirming phase purity of both powder and pressed pellet samples used in this study. As expected, the pressed pellet shows significantly broader Bragg peaks, likely a result of the complex strain texture in these samples which gives a broader distribution of lattice parameters compared to the loose powder samples. A refinement of the powder data produces the $R\overline{3}c$ space group structure with rhombohedral lattice parameters consistent with prior work\cite{Morelock2015,MogusMilankovi1985}.

\section{\label{sec:IR}Infrared Reflectivity} 

To assess the electronic status of TiF$_3$, we seek information on the transport characteristics, which can be challenging in granular samples. Another option besides direct transport measurements is to extrapolate the far-infrared optical response to the static limit, where metals and insulators have strongly contrasting frequency-dependent behavior. Far-infrared reflectivity $R(\omega)$ measurements were conducted using a hydraulically-pressed pellet which was half-coated with gold using a thermal evaporator source. A spectrum taken from the optically thick gold film provides an appropriate reference spectrum which is divided by the sample spectrum to determine the reflectivity. The data were collected using a Bruker 66v Fourier transform infrared spectrometer (FTIR) with a customized chamber suited for near-normal incidence reflectivity. The spectra were collected using a helium-cooled bolometer covering the spectral range 80–1000 cm$^{-1}$ using a layered mylar beam splitter and a silicon carbide globar as a radiation source. To enable these measurements at high temperatures, crossing the cubic-to-tetragonal structural phase transition at $T_s$=340 $K$ and beyond to the high temperatures $T>420K$ where NTE has been reported, we built a special high-temperature infrared reflectivity apparatus and afixed our sample to a silicon carbide heater element using a high temperature ceramic paste. A dc controllable power supply was used to drive the heater element and a K-type thermocouple monitored the local sample temperature during the measurement.

\begin{figure}
\includegraphics[width=3.2 in]{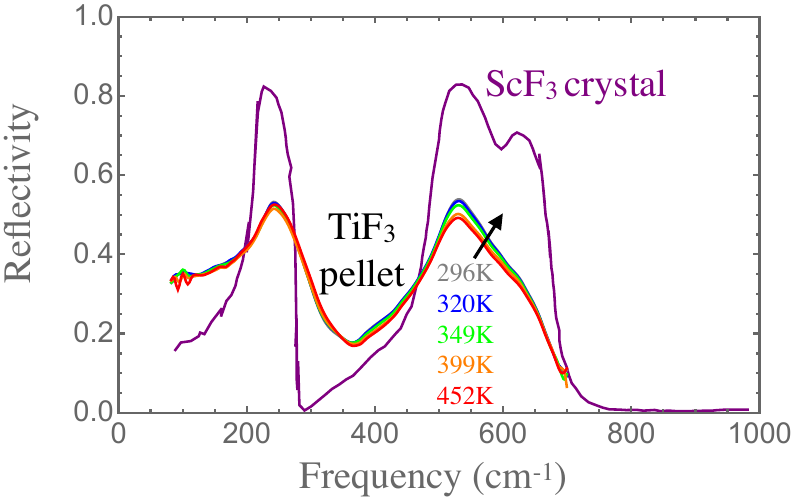}
\caption{Optical reflectivity of packed powder TiF$_3$ at different temperatures crossing from the rhombohedral to cubic structures, where negative thermal expansion has been reported. For comparison, reflectivity of single crystal ScF$_3$ (purple) is shown. Both samples were half coated with a gold film as a reference. Two broad reststrahlen bands show that both systems are insulating.}
\label{fig:TiF3Reflectivity}
\end{figure}

Figure \ref{fig:TiF3Reflectivity} shows the temperature-dependent reflectivity of a pelletized TiF$_3$ sample. In all measured spectra, two strong and broad reststrahlen bands are apparent, very similar to those seen in the single crystal ScF$_3$ reflectivity\cite{Handunkanda2019}, plotted for comparison. We see these features persist when the sample is heated above the structural phase transition and to elevated temperatures where NTE was reported\cite{Morelock2014}. Similar infrared response has been observed in both powdered\cite{Piskunov2016} and single-crystal ScF$_3$ and attributed to the presence of two infrared-active transverse optical phonons\cite{WANG2014}. The relative broadening of the reststrahlen bands in the pelletized sample is likely a result of inhomogeneous broadening due to a complex stress state of the packed powder sample, an effect which is potentially exacerbated by the known pressure-tuned ferroelastic structural transition near room temperature\cite{MogusMilankovi1985,Morelock2015}. 

Importantly, the reflectivity in the dc limit tends toward a constant around $R(\omega\rightarrow 0)\simeq 0.3$ for all temperatures, which is strong evidence that TiF$_3$ is an insulator. If any free carriers are present in the system, no static field could exist in the material bulk, and one consequently expects $R(\omega\rightarrow 0) \rightarrow 1$, which is clearly not the case. This optical determination of insulating behavior is fully consistent with our attempts to measure resistance using a multimeter probe on the same pressed pellet. The low-frequency dielectric constant may also be estimated from the low-frequency reflectance value of $\simeq 0.3$ using Fresnel's equations. We find, using $R(\omega\rightarrow 0)=|(\sqrt{\epsilon_0}-1)/(\sqrt{\epsilon_0}+1)|^2$ that the static dielectric function of TiF$_3$ is $\epsilon_0 \simeq$ 11.7, similar to that of pure silicon, and significantly higher than for ScF$_3$, where $\epsilon_0 \simeq$ 5.8. Since the phonon spectral weight is expected to be very similar, the enhanced polarizability of TiF$_3$ must be attributable to interband transitions at higher frequency, consistent with the Lydanne-Sachs-Teller relation\cite{Ashcroft76,Wooten1972}. Indeed, TiF$_3$ has a gray appearance, while ScF$_3$ is transparent, suggesting a much lower energy electronic threshold in TiF$_3$. We conclude that TiF$_3$ is an insulator in both structural phases of interest and return to the electronic structure below in section \ref{sec:ElectronicStructure}.

\section{\label{sec:MagneticSusceptibility}Magnetic Susceptibility}

\begin{figure}
	\includegraphics[width=3.2in]{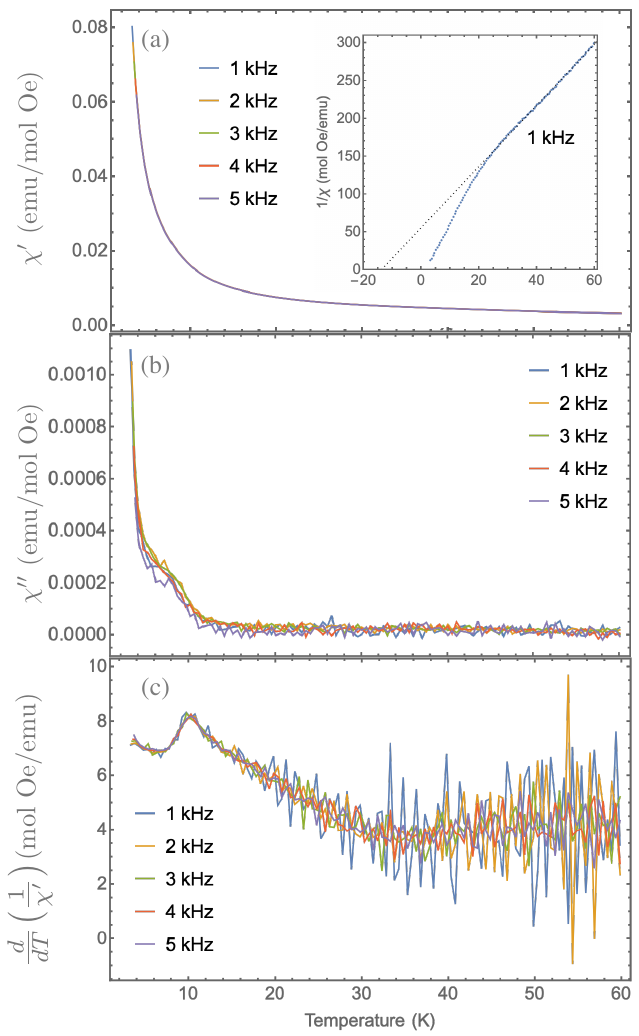}
		
	\caption[]{(a) The in-phase part of the AC susceptibility $\chi'$ taken at 100 Oe from TiF$_3$ at five low frequencies. The inset shows the data for 1kHz plotted as $1/\chi'$ to demonstrate the quality of fit and determination of the antiferromagnetic Weiss temperature $\theta_W$=-13.8K. (b) The out-of-phase AC susceptibility $\chi''$ shows weak features below 10 K. (c) Temperature derivative of the in-phase susceptibility $\frac{d}{dT}\frac{1}{\chi'}$, exposing the features of a magnetic transition below 10K.
}
	\label{fig:TiF3ACsusceptibility}
\end{figure}

Having established that TiF$_3$ shows the infrared response expected of an insulator, we probed its magnetic response to determine whether this is a correlated or band-type insulating system. Magnetic susceptibility measurements were performed in a Quantum Design Physical Property Measurement System. Loose powder was sealed in a nylon capsule in a helium-purged environment for improved thermalization and to mitigate potential oxidation during thermal cycling and material handling. A static field $H$ was applied with a small probing ac field varying at frequency $f$. 
The in-phase ($M'$) and out-of-phase ($M''$) components of the magnetization were measured and the in-phase ($\chi'=M'/H$) and out-of-phase ($\chi''=M''/H$) AC susceptibility was computed. Figure \ref{fig:TiF3ACsusceptibility}a shows AC magnetic susceptibility $\chi'$ with an applied field of 100 Oe varying at frequencies between 1-5 kHz. The Curie-like response appears identical at all measured frequencies, consistent with the presence of local moments in TiF$_3$. The Figure \ref{fig:TiF3ACsusceptibility}a inset shows a plot of $1/\chi'$ for the lowest frequency only along with a fit of data above 20 K to the Curie-Weiss form. Using the high-temperature data $T>30$ K, and performing a least-squares fit to the conventional Curie-Weiss law $\chi(T) = C/(T - \theta_W)$ (dotted line) the Weiss temperature is found to be $\theta_W\simeq$-13.8K with a Curie constant $C$=0.247 emu/K. The Curie constant gives an effective magnetic moment of  $\mu_{eff}$= 1.40 $\mu_B$ per Ti. This value is roughly eighty percent the expected value of 1.73 $\mu_B$ for one spin-1/2 localized electron per Ti$^{+3}$ ion. This local magnetic moment is significantly larger than that of single crystal TiOF which was recently found to be $\mu_{eff}$= 0.95 $\mu_B$ per Ti and a similar Weiss temperature $\theta_W$$\simeq$-19.3 K \cite{Cumby2018}. In that system, a Pauli paramagnetic component was needed to fit the data there and it was suggested that TiOF is partially metallic, or at least near the threshold of electronic delocalization, potentially explaining the significantly lower moment than in TiF$_3$.

Figure \ref{fig:TiF3ACsusceptibility}b shows the out of phase susceptibility $\chi''$, which is consistent with zero for all temperatures above 10 K but shows a steep rise at lower temperature, about 1\% of the in phase part in total. Figure \ref{fig:TiF3ACsusceptibility}c shows the quantity $\frac{d}{dT}(\frac{1}{\chi'})$, which shows clear change of behavior below 10 K. Together, these data suggest a magnetic transition below 10 K. Future neutron magnetic diffraction or resonant X-ray scattering could be used to identify the magnetic state. It is noteworthy that a prior neutron scattering study found an absence of magnetic order down to 10 K\cite{Perebeinos2004}, consistent with the results shown here.

\section{\label{sec:ElectronicStructure}Electronic structure calculations}

We have shown that the optical response of TiF$_3$ is consistent with a correlated insulating state with local moment magnetism. To contextualize these results, we have performed first-principles calculations, both on a full rhombohedral unit cell (space group $R\overline{3}c$, $\#$167) as well as on a symmetry-reduced triclinic cell. In comparison to other work\cite{Perebeinos2004}, we have used a larger unit cell containing 6 Ti and 18 F atoms and also permitted the possibility of non-collinear magnetism. These calculations were carried out within the framework of the density functional theory\cite{a19, a20}, as implemented in the Vienna Ab Initio Simulation Package (VASP) \cite{a22}. The exchange-correlation functional was approximated by the generalized gradient approximation (GGA) \cite{a24}, with the energy cutoff chosen according to potential input files. For integration in $\vec{k}$ - space,  a 6$\times$6$\times$6 (triclinic cell) $\Gamma$ centered mesh according to Monkhorst and Pack \cite{a25} was used during each self-consistent cycle. Several smaller-cell calculations were also run with a 10$\times$10$\times$4 $\Gamma$-centered mesh. Structural optimization was performed until the Hellman-Feynman forces acting on the atoms were negligible. 

\begin{figure}
	\includegraphics[width=3.3 in]{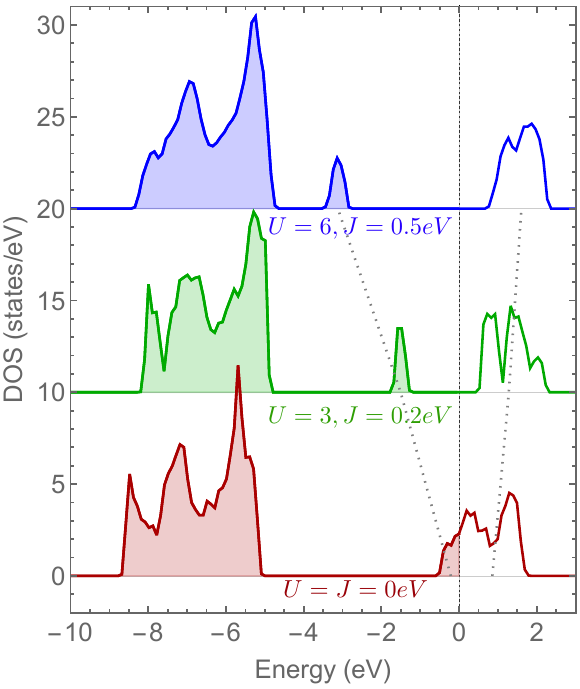}
	\caption{Electronic structure calculations showing the evolution of TiF$_3$ responsive to different values of the correlation parameter $U$ and computed in the GGA+U scheme. The dashed lines track the computed centroid of the unoccupied and occupied $d$ states and are a guide to the eye. $U$=0 corresponds to a ferromagnetic half metal while even modest $U$ values open a sizable correlation gap and favor magnetic order of localized moments. }
	\label{fig:TiF3DOS}
\end{figure}

Figure \ref{fig:TiF3DOS} shows the density of states calculated for different values of the electronic correlation parameter $U$ in the large cell (Ti$_6$F$_{18}$) calculations, which permit freedom for structural optimization compared to prior work. The behavior of the $d$-electrons becomes quite evident from the density of states. In the insulating ($U$$\ge$3 eV) cases shown, there is a well separated density of states peak just below the gap. A count of the total number of electrons in this sharp feature of the density of states yields nearly exactly one $d$ electron per Ti atom, consistent with the measured long moment. Consistent with prior work\cite{Perebeinos2004,Qin2022}, in the uncorrelated $U$=0 eV limit, the calculation produces a finite density of states at the chemical potential $E$=0 for the majority spin species while the electronic states with minority spin are conversely gapped. This structure describes a fully spin polarized ferromagnetic half metal, which is clearly inconsistent with the experimental evidence presented here. In contrast, invocation of even a moderate correlation parameter $U$=3 eV \cite{Yekta2021,esteras2021hubbard,he2016unusual} predicts the development of a gap spanning the chemical potential and an insulating state in the correlated limit. Increasing this parameter to $U$=6 eV further widens this gap, consistent with prior work\cite{Perebeinos2004}. We conclude that correlated electron physics are required in a minimal set of considerations to describe the observed behavior presented here. Interestingly, the degree of correlation required to separate the antiferromagnetic insulating and fully spin-polarized itinerant ferromagnetic half-metal state is rather small. While most calculations are limited exactly at the boundary between these two states\cite{Sachdev2011}, these considerations bolster our assignment of TiF$_3$ as residing at the Mott insulator side of this putative transition and demonstrate the failure of uncorrelated approaches in describing the electronic and magnetic ground state.




\section{\label{sec:Discussion}Discussion}
The realization that TiF$_3$ is a Mott insulator with local moment magnetism severely limits approaches to describing some of its salient properties, such as NTE. For example, a recent work\cite{Qin2022} presents a comparative structural diffraction and pair-distribution analysis of ScF$_3$ and TiF$_3$ data and supplements the predicted ($U$=0) band structure, which predicts a metallic state, similar to what is shown in Fig 4 for $U$=$J$=0. In this framework, the zone boundary Gruneisen parameters of ScF$_3$ are computed to be much more negative than those of TiF$_3$, a difference which is ascribed to screening by the ferromagnetic half metal states, whose existence we show are inconsistent with several of our measurements. In particular, the Gruneisen parameters of both systems are calculated as a function of carrier concentration $n$, and a metallic state of TiF$_3$ is assumed to explain the weaker NTE effect in TiF$_3$. The results we present in Figures 2 and 3 clearly show that $n$=0 for both systems and TiF$_3$ is not at all metallic. Another paper\cite{WANG2014} interprets prior diffraction data\cite{Kennedy2002} as showing NTE at low temperatures and similarly applies a $U$=$J$=0 density functional theory calculation to describe the effect, although the original experimental authors do not in fact claim intrinsic NTE from their data. Still, in light of our conclusion of strong correlations in TiF$_3$, the relative importance of zone-boundary optical phonons on the lower NTE effect in TiF$_3$ relative to ScF$_3$ must be redressed and likely would need to involve correlated electron physics.

In light of the sizable NTE in ScF$_3$ at all temperatures $T<1000K$\cite{Greve2010} and its persistence above $T_s$ in Sc$_{1-x}$Ti$_x$F$_3$\cite{Morelock2014}, it is of interest to consider the nature of the structural phase boundary in Figure \ref{fig:structure}a. On one end, the conventional band insulator ScF$_3$ harbors an incipient structural transition and displays very large and persistent NTE. On the other end, an electronically correlated insulator TiF$_3$ has weaker NTE above 420 K (down to -5ppm/K in one report), and strong positive thermal expansion (up to $\sim$+350ppm/K) below its structural phase transition temperature $T_s$$\sim$340 K with signatures of magnetic order below 10 K. We consider here separately two possible contributors for the structural transition boundary and energy lowering of the distorted structure: (1) a cooperative Jahn-Teller effect and (2) anion dipole-dipole interactions.

In molecular octahedral coordination complexes and crystal systems based on Ti$^{+3}$, a significant Jahn-Teller distortion can be present\cite{Kugel1982}. Briefly, a cubic field splits the free $d$ states into a high energy $e_g$ orbital doublet and a lower energy $t_{2g}$ orbital triplet manifold by a high energy scale 10$D_q$. Introduction of one electron into the lower $t_{2g}$ manifold naively produces a threefold orbital degeneracy per site, setting the circumstances for a Jahn-Teller effect, where any nonzero electron-ion interaction in principle permits a symmetry-lowering relaxation of nuclear coordinates accompanied by an orbital polarization on each site, consistent with the Jahn-Teller theorem. For TiF$_3$, both prior work\cite{Perebeinos2004} and the electronic energy levels in Figure \ref{fig:TiF3DOS} show that a level splitting of this manifold stabilizes a single $a_g$ state polarized along the cubic 111 direction. While the first-order, pseudo-, and second-order Jahn-Teller effects are widely considered to be present in Ti$^{+3}$ crystals\cite{Koppel2009}, their ubiquity in describing structural phase boundaries varies - having been described as negligible and irrelevant for determining ground state properties in some rare-earth titanates\cite{Varignon2017} or requiring additional considerations\cite{Khalsa2018}. In the oxide perovskite LaTiO$_3$ for example, neutron refinement of the magnetic structure\cite{Keimer2000} and advanced theoretical considerations\cite{Khaliullin2000} suggest orbital fluctuations are strong, but are much weaker in YTiO$_3$, suggesting a variety of importance of the Jahn-Teller effect among titanium perovskites\cite{Kugel1982}. In the present case, it is interesting to note that the phase boundary in Figure \ref{fig:structure}a terminates almost exactly at the composition which corresponds to zero $d$ filling, as one may expect for a $d$-electron driven mechanism like the Jahn-Teller distortion. 

On the other hand, while correlated $d$ electron effects can certainly act in the direction of inducing a rhombohedral ground state, there is another possibility suggested by a body of work addressing the $Pm\overline{3}m$ to $R\overline{3}c$ structural transition in AlF$_3$ at an elevated temperature of $T_s$=740 K. Because the occupied states in this system consist only of extended $s$- and $p$-derived electronic orbitals, no electron correlation effects are expected or observed, consistent with the observed diamagnetic susceptibility. This transition has been extensively studied theoretically using density functional theory\cite{Chen2004}, molecular dynamics (MD) simulations\cite{Chaudhuri2004}, and an analytical theory of anion polarizability\cite{Allen2006} which generalizes the Clausius-Mossotti relation\cite{Allen2004}. With the circumstances of a Jahn-Teller distortion completely absent, the driving influences of the transition in AlF$_3$ have been described quantitatively in terms of the long-range electric dipole-dipole interaction of the F$^-$ anion sublattice in the distorted structure. In response to an octahedral tilt pattern relating the cubic and rhombohedral structures, direct anion displacement of the F nuclei give a \textit{direct} nuclear dipole moment $\mu_D$ which is partially compensated by an \textit{induced} electronic polarization $\mu_I$ for the F$^-$ ion. The collective octahedral tilting pattern lowers the energy of the distorted structure significantly and delivers a $R\overline{3}c$ structure in the ground state. In this scenario, electrostatic Madelung energy is the dominant driver of the structural transition. 

Without appealing to a Jahn-Teller mechanism, we may apply the results of the electrostatic theory to TiF$_3$, since ref \cite{Allen2006} describes an analytical theory in terms of a lattice parameter $a$. A surprising result for AlF$_3$ ($a$=3.43$\AA$) is that $\mu_I$$\sim$-0.80$\mu_D$, or in other words, the electronically polarized F$^-$ induced dipole moment opposes the direct one, with about 80\% cancellation of the direct displaced electric dipole. The effect of anion electronic polarizability is therefore to permit a large displacement with only a small change in the overall dipole-dipole electrostatic energy, essentially softening the associated relevant zone-boundary phonon modes considerably. Applying this analytic theory to the present case of TiF$_3$, where $a=3.88\AA$, one obtains $\mu_I$$\sim$-0.49$\mu_D$, suggesting such effects are important in this system as well. While a comparison of transition temperatures between AlF$_3$ and TiF$_3$ is limited in the framework of this theory, the curvature of the expanded potential about the cubic structure of TiF$_3$ is about 62\% of that in AlF$_3$. The observed ratio of transition temperatures is 340 K/740 K$\simeq$46$\%$, which would suggest a curvature ratio $\sqrt{0.46}$$\simeq$0.68. While this analysis is limited, the anion dipole-dipole effects appear to account for most of the changes in the energy landscape in comparing uncorrelated AlF$_3$ and correlated TiF$_3$, without the need to invoke a Jahn-Teller mechanism at all.

Not mentioned in the electrostatic theory, but possibly relevant, is the potential effect of increased dielectric susceptibility due to the presence of interband transitions from $d$-derived bands, which may screen in part the dipole-dipole interactions and reduce the influence of the electrostatic contribution to energy lowering. Extensions of these ideas to transition metal trifluorides, and perhaps a broader class of correlated oxides, would be of value to understanding cooperative effects of Jahn-Teller and dipolar interactions in determining octahedral tilt instabilities in the important class of perovskite-based quantum materials.

In summary, the data presented here show clearly (1) that TiF$_3$ is an insulator, (2) that TiF$_3$ shows local moment spin-$\frac{1}{2}$ magnetism, (3) that TiF$_3$ shows magnetic features below 10 K and (4) that at least a moderate correlation parameter is required to describe the electronic structure of this system. We have suggested that Jahn-Teller effects do not act alone in driving the structural transition and anion polarizability likely plays a significant role in the physics of transition metal trifluorides. Importantly, in these open perovskites NTE materials, electron-correlated effects are present, enhancing the interest and potential applications of NTE, or materials which provide multiple functionalities and controlled thermal expansion.

\begin{acknowledgments}
We acknowledge helpful discussions with Igor Maznichenko, Arthur Ernst, Pavel A. Volkov, and Angus Wilkinson. We are especially grateful to acknowledge support from the U.S. National Science Foundation, award No. NSF-DMR-1905862. R.M.G. acknowledges support from the Swedish Research Council (VR starting Grant No. 2022-03350) and Chalmers University of Technology. We also acknowledge the computing resources provided by the Center for Functional Nanomaterials, which is a U.S. DOE Office of Science Facility, at Brookhaven National Laboratory under Contract No. DE-SC0012704 and the Swedish National Infrastructure for Computing (SNIC) via the National Supercomputer Centre (NSC). 
\end{acknowledgments}

\nocite{*}

\bibliography{tif3bib}

\end{document}